\documentclass[aps,preprint,nofootinbib,superscriptaddress]{revtex4}

\usepackage{amssymb}

\usepackage{epsfig}
   \usepackage{longtable}
\usepackage[bookmarksnumbered,bookmarksopen,colorlinks,citecolor=blue,linkcolor=blue]{hyperref}

\begin{document}

\title{Calculation of cluster decays half-lives for nuclei between $56<Z_p<120$
 by using temperature dependent proximity model  }

\author{V. Zanganeh}
\email{zanganeh\_vahid@yahoo.com ; v.zanganeh@gu.ac.ir}
\affiliation{ Department of Physics, Sciences Faculty, Golestan University, P. O. Box 49138-15759, Gorgan, Iran\\ submitted at : Chinese Physics C
}

\begin{abstract}
Cluster decays half-lives of elements with proton numbers between
$56<Z_p<120$ are calculated by applying temperature dependent
proximity potential approach. for showing the influence of
temperature on cluster decays, we compared the results among
temperature dependent and independent case with experimental values.
The obtained results of the present investigation reveal that we
have more accurate results for temperature dependent proximity
potential in comparison to ignoring one. In the present work, 
we find that results provided with
temperature dependent proximity model are reasonable estimates for
cluster decays half-lives and provide reliable predictions for other
super heavies cluster decays.



\end{abstract}
\maketitle

\begin{center}
\textbf{I. INTRODUCTION}
\end{center}

Spontaneous cluster decays, heavier than alpha particles but lighter
than a fission fragment, of super heavy nuclei is one of most
dominant decay chains which happens before spontaneous fission.
However, the cluster decays and half- lives of the super heavy
nuclei gives us information about the island of stability regions
and hence help us to understand the nuclear structure of the
daughter  as well as parents nuclei\cite{sushil,Poenaru}.
experimentally and theoretically, emitted clusters from heavy nuclei
has greatly attracted researchers attention which theoretical
mechanism of cluster decay is regarded as quantum mechanical
tunneling through the potential barrier between cluster emitter and
the residual daughter nucleus. At this study, calculation of
potential barrier is critical part. At present, many theoretical
approaches have been used to describe the cluster-decay, such as the
macroscopic-microscopic model\cite{Mihail}, Density-dependent
cluster model\cite{Ni:2011zza,Ni:2010zza,Xu:2006tu} relativistic
mean field theory \cite{Bhuyan,Roy}. in these models, various
nuclear potential is used for calculation of half-lives and
spectroscopic properties. The liquid drop model , double-folding
model and proximity are example of most potential that are  applied
recently
\cite{Buck92,Buck93,Xu04,Mohr00,Atzrott96,Xu05,Ren04,Choudhury06}.
One of the successful and applicable models is by using the
proximity potential which is a function of separation between the
surfaces of the two nuclei. Many versions of proximity potential are
proposed by different groups in order to improve the model
\cite{Blocki77,Royer77,shlomo91,Myers00,Dutt2010,Salehi13}.
Interestingly, the temperature dependence of proximity model has
been modified by some authors to study fusion reactions and barrier
characteristic\cite{Salehi13}.

At previous work, we studied the influence of the temperature of the
parent nucleus on the alpha-decay process by applying the
temperature dependence of the proximity potential and transfer
matrix approach to calculate the penetration
probability\cite{vahid2014}. In this work, we attempt to study the
cluster-decay half-lives of parent nuclei by considering
the temperature dependence of the proximity potential and using WKB
approach for calculating the penetration probability across the
potential barrier. The structure of this paper is as follows: In
sec. II, modified proximity model with temperature dependence is
briefly introduced. In sec. III, half-lives of emitted
$^{14}$C,$^{18,20}$O, $^{23}$F, $^{22,24,25,26}$Ne, $^{28,30}$Mg and
$^{34}$Si cluster are compared with existing experimental values. In
addition, half-lives of heavy nuclei are calculated
theoretically and compared with analytical relation based on the ASAF model \cite{Poenaru}.
 Finally the conclusion is given in Sec. IV.

\newpage
\begin{center}
\noindent{\bf {II.  MODEL}}\\
\end{center}

In the cluster model the parent nucleus is assumed to be the
interaction between the cluster particle an daughter nucleus where
the total potential is equal to the sum of the nuclear potential,
the coulomb potential and centrifugal barrier. Thus,

\begin{equation} \label{1}
V(R)=V_N(R)+V_C(R)+\hbar^2 l(l+1)/(2\mu R^2)
\end{equation}
Where $\mu$ is the reduced mass and $V_c(R)$, the Coulomb interaction
potential is given by,

\begin{equation} \label{2}
V_c(R)=\left\{
\begin{array}{ll}
Z_e Z_d e^2/R  & \quad \mbox{for}  R\geq R_c \\
       (Z_e Z_d e^2/2R)(3-(R/R_c)^2)   & \quad \mbox{for}  R\leq R_c
\end{array}
\right.
\end{equation}
in above equation, $R_c$ is expressed by, $R_c=1.24(R_e+R_d)$,$R_e$
and $R_d$ are respectively the radii of emitted cluster and daughter
nuclei. However, $Z_e$ and $Z_d$ represents the charge number of
emitted cluster and daughter nuclei respectively.

Using the proximity theorem we can obtain a simple formula for
nuclear potential between emitted clusters and residual daughter
nuclei as a function of the separation distance between the surfaces
of them \cite{Blocki77}.
\begin{equation} \label{1}
V_N(r)=4\pi \gamma b \overline{R}\Phi(\xi) \quad   MeV.\\
\end{equation}

Here $\overline{R}$ is the reduced radius and is written as:
\begin{equation} \label{2}
\overline{R}=\frac{C_1 C_2}{C_1+C_2}
\end{equation}
and
\begin{equation} \label{3}
C_i=R_i [1-(\frac{b}{R_i})^2+...].
\end{equation}

where b is the surface width and $R_i$ is the effective sharp radius, and given by:
\begin{equation} \label{4}
R_i=1.28A_i^{1/3}-0.76+0.8A_i^{-1/3}  fm \quad\qquad (i=1,2).
\end{equation}
In Eq. (1), $\Phi(\xi)$ is the universal function which has been derived by several authors
in different forms \cite{Myers00,Dutt2010} and in original proximity version was defined as:
\begin{eqnarray} \label{5}
\Phi(\xi)\left\{
\begin{array}{ll}
 -\frac{1}{2} (\xi -2.54)^2-0.0852(\xi-2.54)^3 &  \quad   \xi \leq 1.2511 \\
-3.437 e^{-\xi/0.75} &  \quad \xi> 1.2511
\end{array}
\right.
\end{eqnarray}
and the surface energy coefficient defines as a function of the neutron/proton excess as:
\begin{equation} \label{6}
\gamma=\gamma_0[1-k_s A_s^2]
\end{equation}
where $A_s=(\frac{N-Z}{N+Z})$ and $\gamma_0$ and $K_s$ are the surface energy and surface
asymmetry constants respectively.
These constants have different values in different proximity potential versions
and they revised to $K_s=4$ and $\gamma_0=1.460734 {Mev}/{fm}^2$ for to the proximity-2010 \cite{Dutt2010}
that we used in this work .
In order to achieve an exact form of proximity potential where be able to reproduce
the experimental data more accurately, many researches have been done which
led to different versions for proximity potentials \cite{Winther95, Ngo80, Denisov02}. In one of these attempts,
proximity-2010 is modified with a temperature dependence of surface energy
coefficient and it has been successful in expecting the fusion barrier data and the
experimental fusion cross section \cite{Salehi13}.
 \begin{equation} \label{7}
\gamma(T)=\gamma(T=0)[1-\frac{T-T_B}{T_B}]^{3/2}
\end{equation}
where $T_B$ is the temperature associated with the energies near the Coulomb barrier.\\
Temperature dependency, also followed in some other parts of the proximity potential as:
 \begin{equation} \label{8}
R_i(T)=R_i(T=0)[1+0.0005T^2] fm  (i=1,2)
\end{equation}
and,
\begin{equation} \label{9}
b(T)=b(T=0)[1+0.009T^2]
\end{equation}

 The temperature $T$ in Eqs. (9-11) can be expressed as \cite{Teq1,Teq2},
\begin{equation} \label{10}
E_{CN}^*=E_{c.m.}+Q_{in}=\frac{1}{9}AT^2-T.
\end{equation}
Here, $E_{CN}^*$ denotes the excitation energy of parent nucleus
with mass number $A$. $Q_{in}$ denote the entrance channel Q-value
of the system and $E_{c.m.}$ is the center-of-mass incident energy
which according to Refs. \cite{Salehi13,Golsh13} , one can use the following definition
\begin{equation}
 E_{c.m.}=\frac{e^2 Z_e Z_d}{R_1 +R_2 +2}
\end{equation}
where the radius $R_{1,2}$ is obtained by Eq. (6).
In order to explore the temperature effects of parents nucleus in this study,
we have employed all three above relations simultaneously in proximity-2010
potential, and we have calculated the interaction potential in this way.
With the shape of total cluster-nucleus potential, one can calculate the
penetration probability as well as half-life $T_{1/2}$ of the parent
nucleus. According to the WKB approximation the penetration
probability is calculated by,
\begin{equation} \label{11}
P=\exp\left [-\frac{2}{\hbar}\int_{R_{\rm in}}^{R_{\rm out}}\sqrt{2\mu(R)[V(R)-Q]}\;dR \right ]
\end{equation}
Where $\mu(R)$ is the effective mass of the cluster particle and the
daughter nucleus which is set as the reduced mass. $ Q$ is released energy
for which experimental values are used in the present calculations.
$R_{\rm in}$ and $R_{\rm out}$ denote the
classical turning points inside and outside of the barrier which are
determined from the equation $V(R_{\rm in})=V(R_{\rm out})=Q$.

The cluster-decay half-life $T_{1/2}$ is then calculated with the penetration probability \cite{Thalf},
\begin{equation}
T_{1/2}=\frac{h \ln 2}{2E_\nu P}.
\end{equation}
Where $E_\nu$  denotes the zero point the empirical vibration energy
is given by \cite{Ev},
\begin{equation}
      E_v=Q [0.056+ 0.039  e^{4-A_e\over{2.5}}]
\end{equation}
where $A_e$ is the mass number of emitted cluster nuclei

\begin{center}
\textbf{III. RESULTS}
\end{center}
In this section at first we test our calculation for the existing
measured values of half-lives. after investigate the role of temperature dependence on
cluster decay half-lives, then we will apply this formalism for calculation of
cluster decay half-lives.

\begin{center}
\textbf{III-A. compare to experimental data }
\end{center}

In order to test the precision of our calculation, we compare the
calculated results with the existing experimental data\cite{Royer01}
of the half-lives. We gets the 28 parents nuclei which cluster
decays includes $^{14}$C, $^{18,20}$O, $^{23}$F, $^{22,24,25,26}$Ne,
$^{28,30}$Mg and $^{34}$Si. It is relevant to mention here that the
selected cluster nuclei were discovered from the
experiments\cite{Rose84, Aleksandrov84}. In Table-I, we compared
measured experimental data and the calculated half-lives of cluster
decay with including the temperature effect of parent nucleus(TD.)
and without the temperature effect(IND.). also the minimum angular
momentum $l_{min}$ carried away by the emitted cluster is determined
by the principle of spin-parity conservation when the nuclei are
decayed and the values are from \cite{Basu03}. In order to give some
indication of the quality of the results, the last line of table-I
also shows the relative error,
\begin{equation}
\chi^2_R=\frac{1}{N}\sum_{i=1}^{N}[\frac{Y^{Exp.}-Y^{Th.}}{Y^{Exp.}+Y^{Th.}}]^2,
\end{equation}
where Y=$log_{10}T_{1/2}$.
This quantity show the deviation of the calculated half-lives from the experimental values.
It is clear that corresponding values of $\chi^2_R$ for temperature dependent is less
than ignoring the temperature effects
one.

In order to show better the ability of our method on calculations of cluster decay half-lives,
we define hindrance factor as follow:
\begin{equation}
 HF=\frac{T_{1/2}^{Exp.}}{T_{1/2}^{Cal}}
\end{equation}

Figure-1 depicts the comparison of HF between temperature dependence of proximity potential (TD.) and
temperature independent (IND.). as the definition of HF clear that closer to unity is more accurate results we have.
This figure reveals that TD. results are motivating. therefor it seems we can apply this method for calculation of cluster
decay for heavy nuclei.

\begin{figure}
\includegraphics[angle=0,width=1.0 \textwidth]{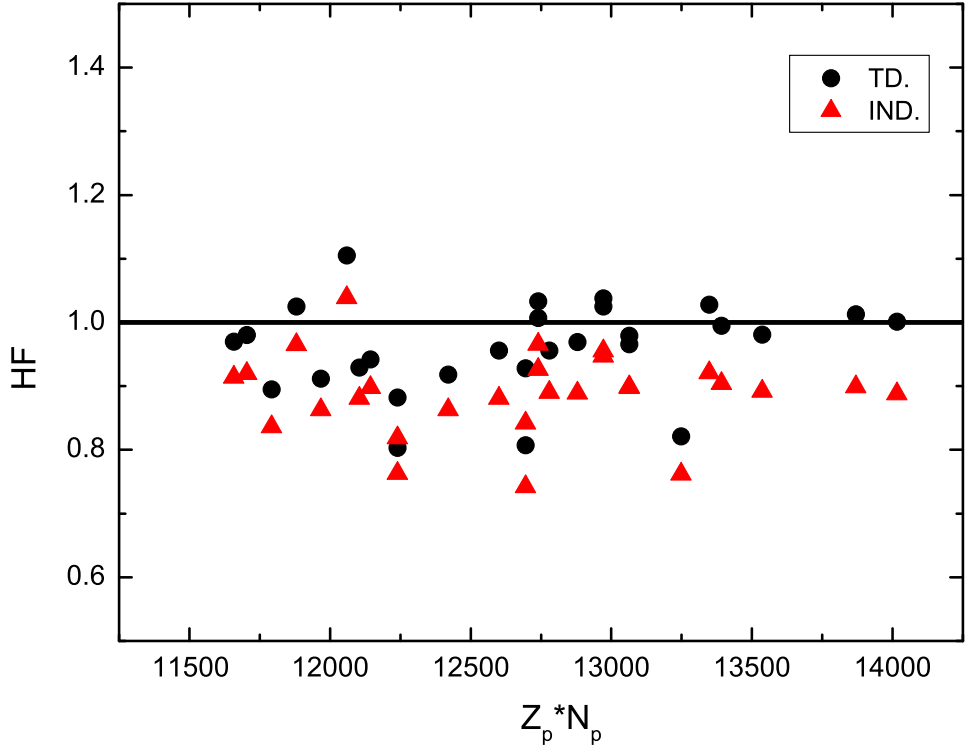}
\caption{Comparison between hindrance factor of temperature
dependence of cluster decays half-lives (TD.) and independent case
(IND.).}
\end{figure}

\begin{center}
\textbf{III-B. Calculation of cluster decay for heavy nuclei}
\end{center}
In Table-II we list the theoretical cluster decay half-lives of
temperature dependent proximity potential  and prediction of ASAF model
\cite{Poenaru} ones for nuclei with Z parent between 56
and 120. The first column of Table-II denotes nuclide. The
experimental Q cluster-decay energy ($Q_c$) is given in column 2.
When the $Q_c$ is not known, we use a theoretical value from ref.
\cite{ning}. The logarithmic cluster decay half-lives with TD. and
ASAF model \cite{Poenaru} ones are listed in columns 3 and 4,
respectively. The last column is the reference the error between
calculation of our work and ASAF model. It is seen from Table-II
that TD. half-lives are reasonable estimates.\\

\newpage
\begin{center}
\textbf{IV. CONCLUSION}
\end{center}
By using the temperature dependent proximity potential, we calculated
cluster decay half-lives  of parent nuclei whose proton numbers are
from $Z_p=56$ to $Z_p=120$. Before the calculation,
 we test the accuracy of the our calculation for some nuclei
in compare with experimental data. The results of present
calculation made with T.D. model  are in good agreement with
experimental data (see Table-I and Figure-1). at next, T.D. model
calculation are provided in Table-II for cluster decay half-lives of
heavy nuclei and compared with the values based on the ASAF model
\cite{Poenaru} estimation using the same $Q-values$. this formalism
has been found to be quit reliable.

\begin{center}
 \textbf{ACKNOWLEDGMENTS}
\end{center}
The authors would like to give special thanks to Dr. F. Zanganeh for
helpful discussions and encouragements.

\newpage
\noindent{\bf {TABLE CAPTIONS}}
\\
\\

\footnotesize {
\noindent{Table-I: Comparison of logarithmic half-lives of cluster decays between temperature independent proximity
(${IND.}$),temperature dependent proximity (${TD.}$) and experiment data}\\
\vskip 0.4cm
\begin{tabular}{|l|lllllll|}
\hline
cluster decay&  lmin \quad Q$_{exp.}$ \quad ${IND.}$ \qquad ${TD.}$\qquad
${Exp.}$ \qquad $\Delta^{IND.}$ \quad $\Delta^{TD.}$&\\
\hline
$^{221}$Fr $\longrightarrow$ $^{14}$C &   3 \quad  31.29  \quad \quad  15.86 \quad \quad  14.94 \quad \quad  14.5 \quad \quad   -1.36  \quad   -0.44& \\
$^{221}$Ra $\longrightarrow$ $^{14}$C &   3 \quad  32.40  \quad \quad  14.56 \quad \quad  13.66 \quad \quad  13.4 \quad \quad   -1.16  \quad   -0.26& \\
$^{222}$Ra $\longrightarrow$ $^{14}$C &   0 \quad  33.05  \quad \quad  13.15 \quad \quad  12.28 \quad \quad  11.0 \quad \quad   -2.15  \quad   -1.28& \\
$^{223}$Ra $\longrightarrow$ $^{14}$C &   4 \quad  31.83  \quad \quad  15.75 \quad \quad  14.82 \quad \quad  15.2 \quad \quad   -0.55  \quad    0.37&  \\
$^{224}$Ra $\longrightarrow$ $^{14}$C &   0 \quad  30.54  \quad \quad  18.19 \quad \quad  17.22 \quad \quad  15.7 \quad \quad   -2.49  \quad   -1.52& \\
$^{226}$Ra $\longrightarrow$ $^{14}$C &   0 \quad  28.19  \quad \quad  23.61 \quad \quad  22.51 \quad \quad  21.2 \quad \quad   -2.41  \quad   -1.31& \\
$^{225}$Ac $\longrightarrow$ $^{14}$C &   4 \quad  30.47  \quad \quad  19.52 \quad \quad  18.52 \quad \quad  17.2 \quad \quad   -2.32  \quad   -1.32& \\
$^{226}$Th $\longrightarrow$ $^{14}$C &   0 \quad  30.55  \quad \quad  20.04 \quad \quad  19.04 \quad \quad  15.3 \quad \quad   -4.74  \quad   -3.74& \\
$^{226}$Th $\longrightarrow$ $^{18}$O &   0 \quad  45.73  \quad \quad  20.50 \quad \quad  19.05 \quad \quad  16.8 \quad \quad   -3.70  \quad   -2.25& \\
$^{224}$Th $\longrightarrow$ $^{14}$C &   0 \quad  32.93  \quad \quad  15.11 \quad \quad  14.20 \quad \quad  15.7 \quad \quad    0.58  \quad    1.49&  \\
$^{228}$Th $\longrightarrow$ $^{20}$O &   0 \quad  44.73  \quad \quad  23.97 \quad \quad  22.53 \quad \quad  20.7 \quad \quad   -3.27  \quad   -1.83& \\
$^{231}$Pa $\longrightarrow$ $^{23}$F &   1 \quad  51.85  \quad \quad  26.90 \quad \quad  25.16 \quad \quad  26.0 \quad \quad   -0.90  \quad    0.83&  \\
$^{230}$U  $\longrightarrow$ $^{22}$Ne&   0 \quad  61.39  \quad \quad  23.27 \quad \quad  21.11 \quad \quad  19.6 \quad \quad   -3.67  \quad   -1.51& \\
$^{230}$Th $\longrightarrow$ $^{24}$Ne&   0 \quad  57.76  \quad \quad  27.92 \quad \quad  25.72 \quad \quad  24.6 \quad \quad   -3.32  \quad   -1.12& \\
$^{231}$Pa $\longrightarrow$ $^{24}$Ne&   1 \quad  60.41  \quad \quad  24.71 \quad \quad  22.73 \quad \quad  22.9 \quad \quad   -1.81  \quad    0.16&  \\
$^{230}$U  $\longrightarrow$ $^{24}$Ne&   0 \quad  61.35  \quad \quad  24.51 \quad \quad  22.56 \quad \quad  18.2 \quad \quad   -6.31  \quad   -4.36& \\
$^{232}$U  $\longrightarrow$ $^{24}$Ne&   0 \quad  62.31  \quad \quad  22.94 \quad \quad  21.05 \quad \quad  20.4 \quad \quad   -2.54  \quad   -0.65& \\
$^{233}$U  $\longrightarrow$ $^{24}$Ne&   2 \quad  60.49  \quad \quad  25.97 \quad \quad  23.89 \quad \quad  24.8 \quad \quad   -1.17  \quad    0.90&  \\
$^{234}$U  $\longrightarrow$ $^{24}$Ne&   0 \quad  58.83  \quad \quad  28.91 \quad \quad  26.55 \quad \quad  26.0 \quad \quad   -2.91  \quad   -0.55& \\
$^{233}$U  $\longrightarrow$ $^{25}$Ne&   2 \quad  60.73  \quad \quad  26.17 \quad \quad  24.19 \quad \quad  24.8 \quad \quad   -1.37  \quad    0.60&  \\
$^{232}$Th $\longrightarrow$ $^{26}$Ne&   0 \quad  55.97  \quad \quad  32.56 \quad \quad  30.32 \quad \quad  29.0 \quad \quad   -3.56  \quad   -1.32& \\
$^{234}$U  $\longrightarrow$ $^{26}$Ne&   0 \quad  59.47  \quad \quad  28.94 \quad \quad  26.91 \quad \quad  26.0 \quad \quad   -2.94  \quad   -0.91& \\
$^{236}$U  $\longrightarrow$ $^{26}$Ne&   0 \quad  56.75  \quad \quad  34.10 \quad \quad  31.67 \quad \quad  26.0 \quad \quad   -8.10  \quad   -5.67& \\
\hline
\end{tabular}

\begin{tabular}{|l|lllllll|}
\hline
cluster decay&  lmin \quad Q$_{exp.}$ \quad ${IND.}$ \qquad ${TD.}$\qquad
${Exp.}$ \qquad $\Delta^{IND.}$ \quad $\Delta^{TD.}$&\\
\hline

$^{236}$Pu $\longrightarrow$ $^{28}$Mg&   0 \quad  79.67  \quad \quad  23.55 \quad \quad  21.10 \quad \quad  21.7 \quad \quad   -1.85  \quad    0.59&  \\
$^{237}$Np $\longrightarrow$ $^{30}$Mg&   2 \quad  74.79  \quad \quad  30.53 \quad \quad  27.74 \quad \quad  27.6 \quad \quad   -2.93  \quad   -0.14& \\
$^{238}$Pu $\longrightarrow$ $^{30}$Mg&   0 \quad  76.80  \quad \quad  28.81 \quad \quad  26.19 \quad \quad  25.7 \quad \quad   -3.11  \quad   -0.49& \\
$^{241}$Am $\longrightarrow$ $^{34}$Si&   3 \quad  93.93  \quad \quad  28.13 \quad \quad  24.97 \quad \quad  25.3 \quad \quad   -2.83  \quad    0.32&  \\
$^{242}$Cm $\longrightarrow$ $^{34}$Si&   0 \quad  96.52  \quad \quad  26.13 \quad \quad  23.17 \quad \quad  23.2 \quad \quad   -2.93  \quad    0.02& \\
\hline
\textcolor{red}{\emph{$\chi^2_R*10^{-3}$ }} & \textcolor{red}{----}
\quad \textcolor{red}{----} \quad \quad   \textcolor{red}{4.96}
 \quad \qquad   \textcolor{red}{1.91} \quad \qquad \textcolor{red}{----}
 \quad \quad  \textcolor{red}{----} \quad \quad  \textcolor{red}{----} & \\
\hline
\end{tabular}

\newpage
\noindent{Table-II: Comparison of the calculated logarithmic half-lives between temperature dependent proximity (TD.)
with results based on ASAF model \cite{Poenaru}.}  \\
\\

\begin{tabular}{|lll|   |llll|}
\hline
cluster decay &\quad\quad ASAF \quad TD. \qquad $\Delta$ && cluster decay &\quad\quad ASAF \quad TD. \qquad $\Delta$\\
\hline
$^{114}$Ba $\longrightarrow$ $^{12}$C &\quad\quad10.760  \quad10.714\quad-0.046  &&$^{232}$Pa $\longrightarrow$ $^{23}$F &\quad\quad27.041  \quad28.081\quad 1.040  \\
$^{117}$Ba $\longrightarrow$ $^{12}$C &\quad\quad21.159  \quad21.623\quad 0.464  &&$^{232}$U  $\longrightarrow$ $^{26}$Ne&\quad\quad28.735  \quad29.535\quad 0.800  \\
$^{119}$Ba $\longrightarrow$ $^{12}$C &\quad\quad24.044  \quad24.624\quad 0.580  &&$^{232}$Pu $\longrightarrow$ $^{22}$Ne&\quad\quad21.116  \quad22.124\quad 1.008  \\
$^{120}$La $\longrightarrow$ $^{12}$C &\quad\quad23.668  \quad24.256\quad 0.587  &&$^{233}$U  $\longrightarrow$ $^{25}$Ne&\quad\quad23.707  \quad24.162\quad 0.455  \\
$^{121}$La $\longrightarrow$ $^{12}$C &\quad\quad27.208  \quad27.940\quad 0.732  &&$^{233}$U  $\longrightarrow$ $^{28}$Mg&\quad\quad25.106  \quad25.918\quad 0.813  \\
$^{122}$Ce $\longrightarrow$ $^{12}$C &\quad\quad26.027  \quad26.741\quad 0.714  &&$^{233}$Np $\longrightarrow$ $^{24}$Ne&\quad\quad22.049  \quad22.519\quad 0.470  \\
$^{124}$Ce $\longrightarrow$ $^{12}$C &\quad\quad34.286  \quad35.304\quad 1.018  &&$^{233}$Pu $\longrightarrow$ $^{22}$Ne&\quad\quad23.483  \quad24.967\quad 1.484  \\
$^{215}$At $\longrightarrow$ $^{ 8}$Be&\quad\quad15.358  \quad16.694\quad 1.336  &&$^{234}$U  $\longrightarrow$ $^{24}$Ne&\quad\quad25.367  \quad26.550\quad 1.183  \\
$^{218}$Ra $\longrightarrow$ $^{12}$C &\quad\quad14.007  \quad15.194\quad 1.186  &&$^{234}$U  $\longrightarrow$ $^{28}$Mg&\quad\quad25.176  \quad26.074\quad 0.898  \\
$^{222}$Ac $\longrightarrow$ $^{12}$C &\quad\quad12.934  \quad14.093\quad 1.159  &&$^{234}$Np $\longrightarrow$ $^{28}$Mg&\quad\quad22.795  \quad23.059\quad 0.264  \\
$^{223}$Th $\longrightarrow$ $^{15}$N &\quad\quad16.878  \quad17.940\quad 1.062  &&$^{234}$Pu $\longrightarrow$ $^{28}$Mg&\quad\quad21.889  \quad21.856\quad-0.033  \\
$^{225}$Np $\longrightarrow$ $^{12}$C &\quad\quad 9.911  \quad10.645\quad 0.734  &&$^{235}$U  $\longrightarrow$ $^{25}$Ne&\quad\quad27.945  \quad29.198\quad 1.254  \\
$^{225}$U  $\longrightarrow$ $^{15}$N &\quad\quad16.845  \quad17.857\quad 1.012  &&$^{235}$U  $\longrightarrow$ $^{28}$Mg&\quad\quad27.560  \quad29.050\quad 1.490  \\
$^{227}$Th $\longrightarrow$ $^{22}$Ne&\quad\quad25.220  \quad27.049\quad 1.829  &&$^{235}$Np $\longrightarrow$ $^{28}$Mg&\quad\quad22.879  \quad23.228\quad 0.349  \\
$^{228}$Pu $\longrightarrow$ $^{15}$N &\quad\quad18.105  \quad19.258\quad 1.152  &&$^{235}$Pu $\longrightarrow$ $^{25}$Ne&\quad\quad26.781  \quad27.633\quad 0.852  \\
$^{228}$Th $\longrightarrow$ $^{24}$Ne&\quad\quad25.307  \quad26.338\quad 1.031  &&$^{235}$Pu $\longrightarrow$ $^{29}$Mg&\quad\quad25.538  \quad25.911\quad 0.373  \\
$^{228}$Pa $\longrightarrow$ $^{22}$Ne&\quad\quad22.038  \quad23.266\quad 1.228  &&$^{236}$U  $\longrightarrow$ $^{26}$Ne&\quad\quad30.376  \quad31.677\quad 1.301  \\
$^{229}$Ac $\longrightarrow$ $^{22}$O &\quad\quad26.916  \quad27.595\quad 0.679  &&$^{114}$Ba $\longrightarrow$ $^{16}$O &\quad\quad15.192  \quad14.988\quad-0.204  \\
$^{229}$Th $\longrightarrow$ $^{24}$Ne&\quad\quad24.642  \quad25.623\quad 0.981  &&$^{117}$Ba $\longrightarrow$ $^{16}$O &\quad\quad24.113  \quad24.468\quad 0.355  \\
$^{229}$Pa $\longrightarrow$ $^{23}$F &\quad\quad27.024  \quad27.898\quad 0.875  &&$^{119}$Ba $\longrightarrow$ $^{16}$O &\quad\quad27.517  \quad28.064\quad 0.548  \\
$^{229}$Pa $\longrightarrow$ $^{24}$Ne&\quad\quad23.278  \quad23.935\quad 0.656  &&$^{120}$Ce $\longrightarrow$ $^{16}$O &\quad\quad17.415  \quad17.388\quad-0.027  \\
$^{230}$Th $\longrightarrow$ $^{23}$F &\quad\quad28.734  \quad29.995\quad 1.261  &&$^{121}$Ce $\longrightarrow$ $^{16}$O &\quad\quad19.671  \quad19.799\quad 0.129  \\
$^{230}$Pa $\longrightarrow$ $^{23}$F &\quad\quad25.436  \quad26.150\quad 0.714  &&$^{123}$Ce $\longrightarrow$ $^{16}$O &\quad\quad24.949  \quad25.413\quad 0.464  \\
$^{230}$Pa $\longrightarrow$ $^{22}$Ne&\quad\quad25.066  \quad26.958\quad 1.892  &&$^{124}$Pr $\longrightarrow$ $^{16}$O &\quad\quad24.196  \quad24.645\quad 0.449  \\
$^{230}$U  $\longrightarrow$ $^{24}$Ne&\quad\quad22.139  \quad22.566\quad 0.428  &&$^{218}$Fr $\longrightarrow$ $^{ 8}$Be&\quad\quad10.856  \quad11.924\quad 1.067  \\
$^{231}$U  $\longrightarrow$ $^{22}$Ne&\quad\quad22.688  \quad24.111\quad 1.423  &&$^{221}$Pa $\longrightarrow$ $^{ 8}$Be&\quad\quad 8.441  \quad 9.279\quad 0.838  \\

\hline
\end{tabular}
\newpage
\begin{tabular}{|lll|   |llll|}
\hline
cluster decay &\quad\quad ASAF \quad TD. \qquad $\Delta$ && cluster decay &\quad\quad ASAF \quad TD. \qquad $\Delta$\\
\hline
$^{231}$Th $\longrightarrow$ $^{25}$Ne&\quad\quad26.970  \quad28.056\quad 1.086  &&$^{222}$Ac $\longrightarrow$ $^{15}$N &\quad\quad14.665  \quad15.497\quad 0.832  \\
$^{231}$Pa $\longrightarrow$ $^{23}$F &\quad\quad24.509  \quad25.152\quad 0.643  &&$^{223}$Th $\longrightarrow$ $^{17}$O &\quad\quad19.915  \quad21.289\quad 1.374  \\
$^{231}$Np $\longrightarrow$ $^{22}$Ne&\quad\quad20.605  \quad21.555\quad 0.950  &&$^{225}$Pa $\longrightarrow$ $^{15}$N &\quad\quad14.445  \quad15.235\quad 0.789  \\
$^{227}$U  $\longrightarrow$ $^{17}$O &\quad\quad18.959  \quad20.233\quad 1.274  &&$^{235}$U  $\longrightarrow$ $^{29}$Mg&\quad\quad27.959  \quad29.076\quad 1.117  \\
$^{227}$Pa $\longrightarrow$ $^{22}$Ne&\quad\quad22.813  \quad24.119\quad 1.306  &&$^{235}$Np $\longrightarrow$ $^{29}$Mg&\quad\quad27.498  \quad28.386\quad 0.887  \\
$^{228}$Th $\longrightarrow$ $^{22}$Ne&\quad\quad25.723  \quad27.714\quad 1.991  &&$^{235}$Pu $\longrightarrow$ $^{28}$Mg&\quad\quad21.263  \quad21.151\quad-0.112  \\
$^{228}$Ac $\longrightarrow$ $^{23}$F &\quad\quad28.811  \quad30.074\quad 1.263  &&$^{236}$U  $\longrightarrow$ $^{24}$Ne&\quad\quad29.604  \quad31.640\quad 2.036  \\
$^{228}$U  $\longrightarrow$ $^{22}$Ne&\quad\quad20.768  \quad21.681\quad 0.912  &&$^{236}$U  $\longrightarrow$ $^{30}$Mg&\quad\quad29.083  \quad30.084\quad 1.001  \\
$^{229}$Ac $\longrightarrow$ $^{23}$F &\quad\quad27.984  \quad29.195\quad 1.211  &&$^{236}$Np $\longrightarrow$ $^{28}$Mg&\quad\quad25.100  \quad25.992\quad 0.891  \\
$^{229}$U  $\longrightarrow$ $^{22}$Ne&\quad\quad19.874  \quad20.682\quad 0.809  &&$^{236}$Np $\longrightarrow$ $^{30}$Mg&\quad\quad27.581  \quad28.140\quad 0.559  \\
$^{229}$Pa $\longrightarrow$ $^{22}$Ne&\quad\quad22.306  \quad23.642\quad 1.336  &&$^{236}$Pu $\longrightarrow$ $^{29}$Mg&\quad\quad26.214  \quad26.785\quad 0.572  \\
$^{230}$Th $\longrightarrow$ $^{22}$O &\quad\quad26.388  \quad26.962\quad 0.575  &&$^{237}$Np $\longrightarrow$ $^{30}$Mg&\quad\quad27.163  \quad27.708\quad 0.545  \\
$^{230}$Th $\longrightarrow$ $^{24}$Ne&\quad\quad24.674  \quad25.724\quad 1.050  &&$^{237}$Pu $\longrightarrow$ $^{29}$Mg&\quad\quad24.362  \quad24.614\quad 0.252  \\
$^{230}$Pa $\longrightarrow$ $^{24}$Ne&\quad\quad22.249  \quad22.793\quad 0.544  &&$^{237}$Pu $\longrightarrow$ $^{32}$Si&\quad\quad25.273  \quad25.624\quad 0.351  \\
$^{230}$U  $\longrightarrow$ $^{20}$O &\quad\quad25.653  \quad26.672\quad 1.019  &&$^{237}$Am $\longrightarrow$ $^{29}$Mg&\quad\quad27.384  \quad28.142\quad 0.758  \\
$^{231}$Pa $\longrightarrow$ $^{22}$O &\quad\quad29.269  \quad30.111\quad 0.842  &&$^{238}$Pu $\longrightarrow$ $^{28}$Mg&\quad\quad25.341  \quad26.288\quad 0.947  \\
$^{231}$Th $\longrightarrow$ $^{24}$Ne&\quad\quad27.253  \quad28.795\quad 1.542  &&$^{238}$Pu $\longrightarrow$ $^{30}$Mg&\quad\quad25.954  \quad26.195\quad 0.240  \\
$^{231}$U  $\longrightarrow$ $^{22}$O &\quad\quad33.553  \quad34.746\quad 1.193  &&$^{238}$Pu $\longrightarrow$ $^{33}$Si&\quad\quad28.717  \quad29.571\quad 0.854  \\
$^{231}$Pa $\longrightarrow$ $^{24}$Ne&\quad\quad22.144  \quad22.722\quad 0.579  &&$^{238}$Am $\longrightarrow$ $^{29}$Mg&\quad\quad25.778  \quad26.267\quad 0.490  \\
$^{232}$Th $\longrightarrow$ $^{26}$Ne&\quad\quad29.211  \quad30.325\quad 1.114  &&$^{238}$Am $\longrightarrow$ $^{32}$Si&\quad\quad23.191  \quad22.941\quad-0.250  \\
$^{232}$U  $\longrightarrow$ $^{24}$Ne&\quad\quad20.755  \quad21.050\quad 0.295  &&$^{238}$Cm $\longrightarrow$ $^{28}$Mg&\quad\quad22.733  \quad22.893\quad 0.159  \\
$^{232}$U  $\longrightarrow$ $^{28}$Mg&\quad\quad25.065  \quad25.804\quad 0.739  &&$^{239}$Pu $\longrightarrow$ $^{30}$Mg&\quad\quad27.997  \quad28.712\quad 0.715  \\
$^{233}$U  $\longrightarrow$ $^{24}$Ne&\quad\quad23.106  \quad23.847\quad 0.741  &&$^{239}$Pu $\longrightarrow$ $^{34}$Si&\quad\quad27.318  \quad27.455\quad 0.137  \\
$^{233}$U  $\longrightarrow$ $^{26}$Ne&\quad\quad27.197  \quad27.825\quad 0.628  &&$^{239}$Am $\longrightarrow$ $^{32}$Si&\quad\quad23.367  \quad23.227\quad-0.140  \\
$^{233}$Np $\longrightarrow$ $^{22}$Ne&\quad\quad26.062  \quad28.111\quad 2.049  &&$^{239}$Am $\longrightarrow$ $^{34}$Si&\quad\quad26.224  \quad25.957\quad-0.267  \\
$^{233}$Np $\longrightarrow$ $^{25}$Ne&\quad\quad27.596  \quad28.557\quad 0.961  &&$^{240}$Pu $\longrightarrow$ $^{34}$Si&\quad\quad27.029  \quad27.170\quad 0.141  \\
$^{233}$Pu $\longrightarrow$ $^{24}$Ne&\quad\quad22.945  \quad23.473\quad 0.527  &&$^{240}$Am $\longrightarrow$ $^{34}$Si&\quad\quad25.576  \quad25.215\quad-0.361  \\
$^{234}$U  $\longrightarrow$ $^{26}$Ne&\quad\quad26.359  \quad26.914\quad 0.555  &&$^{240}$Cm $\longrightarrow$ $^{32}$Si&\quad\quad21.634  \quad20.979\quad-0.655  \\
$^{234}$Np $\longrightarrow$ $^{25}$Ne&\quad\quad25.498  \quad26.186\quad 0.688  &&$^{241}$Am $\longrightarrow$ $^{33}$Si&\quad\quad28.547  \quad29.451\quad 0.904  \\
$^{234}$Pu $\longrightarrow$ $^{24}$Ne&\quad\quad23.037  \quad23.629\quad 0.592  &&$^{241}$Cm $\longrightarrow$ $^{32}$Si&\quad\quad23.649  \quad23.592\quad-0.058  \\
\hline
\end{tabular}
\newpage
\begin{tabular}{|lll|   |llll|}
\hline
cluster decay &\quad\quad ASAF \quad TD. \qquad $\Delta$ && cluster decay &\quad\quad ASAF \quad TD. \qquad $\Delta$\\
\hline
$^{235}$U  $\longrightarrow$ $^{24}$Ne&\quad\quad27.478  \quad29.079\quad 1.601  &&$^{242}$Cm $\longrightarrow$ $^{34}$Si&\quad\quad23.938  \quad23.174\quad-0.764  \\
$^{235}$U  $\longrightarrow$ $^{26}$Ne&\quad\quad28.323  \quad29.244\quad 0.921  &&$^{242}$Cf $\longrightarrow$ $^{33}$Si&\quad\quad26.205  \quad26.179\quad-0.026  \\
$^{242}$Cf $\longrightarrow$ $^{36}$S &\quad\quad24.037  \quad23.064\quad-0.973  &&$^{237}$Am $\longrightarrow$ $^{28}$Mg&\quad\quad22.160  \quad22.243\quad 0.083  \\
$^{244}$Cm $\longrightarrow$ $^{34}$Si&\quad\quad27.261  \quad27.461\quad 0.200  &&$^{237}$Am $\longrightarrow$ $^{32}$Si&\quad\quad23.567  \quad23.352\quad-0.215  \\
$^{244}$Cf $\longrightarrow$ $^{36}$S &\quad\quad23.781  \quad22.852\quad-0.929  &&$^{238}$Pu $\longrightarrow$ $^{29}$Mg&\quad\quad28.028  \quad29.105\quad 1.077  \\
$^{249}$Cf $\longrightarrow$ $^{42}$S &\quad\quad31.708  \quad30.886\quad-0.822  &&$^{238}$Pu $\longrightarrow$ $^{32}$Si&\quad\quad25.484  \quad25.966\quad 0.482  \\
$^{249}$No $\longrightarrow$ $^{48}$Ca&\quad\quad27.237  \quad23.322\quad-3.915  &&$^{238}$Am $\longrightarrow$ $^{28}$Mg&\quad\quad23.892  \quad24.410\quad 0.518  \\
$^{250}$No $\longrightarrow$ $^{48}$Ca&\quad\quad26.894  \quad22.916\quad-3.979  &&$^{238}$Am $\longrightarrow$ $^{30}$Mg&\quad\quad28.052  \quad28.592\quad 0.540  \\
$^{251}$No $\longrightarrow$ $^{48}$Ca&\quad\quad26.646  \quad22.633\quad-4.013  &&$^{238}$Am $\longrightarrow$ $^{33}$Si&\quad\quad26.006  \quad26.035\quad 0.029  \\
$^{252}$Cf $\longrightarrow$ $^{50}$Ca&\quad\quad32.209  \quad30.055\quad-2.154  &&$^{238}$Cm $\longrightarrow$ $^{32}$Si&\quad\quad22.065  \quad21.411\quad-0.655  \\
$^{252}$No $\longrightarrow$ $^{48}$Ca&\quad\quad26.321  \quad22.250\quad-4.071  &&$^{239}$Pu $\longrightarrow$ $^{33}$Si&\quad\quad27.436  \quad28.043\quad 0.608  \\
$^{253}$No $\longrightarrow$ $^{48}$Ca&\quad\quad26.069  \quad21.964\quad-4.105  &&$^{239}$Am $\longrightarrow$ $^{30}$Mg&\quad\quad27.533  \quad28.027\quad 0.494  \\
$^{253}$Fm $\longrightarrow$ $^{48}$Ca&\quad\quad27.932  \quad24.796\quad-3.137  &&$^{239}$Am $\longrightarrow$ $^{33}$Si&\quad\quad26.563  \quad26.798\quad 0.236  \\
$^{254}$No $\longrightarrow$ $^{46}$Ca&\quad\quad26.810  \quad23.880\quad-2.930  &&$^{239}$Cm $\longrightarrow$ $^{32}$Si&\quad\quad21.546  \quad20.818\quad-0.728  \\
$^{254}$No $\longrightarrow$ $^{50}$Ca&\quad\quad29.401  \quad25.731\quad-3.670  &&$^{240}$Am $\longrightarrow$ $^{33}$Si&\quad\quad25.545  \quad25.602\quad 0.056  \\
$^{255}$No $\longrightarrow$ $^{50}$Ca&\quad\quad28.718  \quad24.871\quad-3.846  &&$^{240}$Cm $\longrightarrow$ $^{30}$Mg&\quad\quad28.773  \quad29.466\quad 0.693  \\
$^{256}$No $\longrightarrow$ $^{50}$Ca&\quad\quad27.896  \quad23.813\quad-4.083  &&$^{240}$Cm $\longrightarrow$ $^{34}$Si&\quad\quad25.142  \quad24.544\quad-0.598  \\
$^{257}$No $\longrightarrow$ $^{50}$Ca&\quad\quad27.034  \quad22.714\quad-4.320  &&$^{241}$Am $\longrightarrow$ $^{34}$Si&\quad\quad25.285  \quad24.923\quad-0.362  \\
$^{258}$No $\longrightarrow$ $^{48}$Ca&\quad\quad27.071  \quad23.640\quad-3.431  &&$^{242}$Cm $\longrightarrow$ $^{32}$Si&\quad\quad25.389  \quad25.858\quad 0.469  \\
$^{258}$Rf $\longrightarrow$ $^{48}$Ca&\quad\quad25.519  \quad21.218\quad-4.301  &&$^{242}$Cf $\longrightarrow$ $^{32}$Si&\quad\quad22.321  \quad21.737\quad-0.584  \\
$^{258}$Rf $\longrightarrow$ $^{50}$Ca&\quad\quad28.953  \quad25.076\quad-3.876  &&$^{242}$Cf $\longrightarrow$ $^{34}$Si&\quad\quad26.373  \quad25.969\quad-0.404  \\
$^{258}$Rf $\longrightarrow$ $^{51}$Ti&\quad\quad27.699  \quad23.286\quad-4.413  &&$^{243}$Cm $\longrightarrow$ $^{34}$Si&\quad\quad25.638  \quad25.360\quad-0.278  \\
$^{258}$Rf $\longrightarrow$ $^{53}$Ti&\quad\quad29.963  \quad25.593\quad-4.370  &&$^{244}$Cf $\longrightarrow$ $^{34}$Si&\quad\quad25.575  \quad25.095\quad-0.480  \\
$^{259}$No $\longrightarrow$ $^{48}$Ca&\quad\quad28.184  \quad25.229\quad-2.955  &&$^{246}$Cf $\longrightarrow$ $^{38}$S &\quad\quad26.050  \quad25.058\quad-0.991  \\
$^{260}$No $\longrightarrow$ $^{48}$Ca&\quad\quad29.194  \quad26.700\quad-2.493  &&$^{249}$Cf $\longrightarrow$ $^{46}$Ar&\quad\quad31.476  \quad29.809\quad-1.668  \\
$^{261}$No $\longrightarrow$ $^{50}$Ca&\quad\quad29.409  \quad26.204\quad-3.205  &&$^{249}$Cf $\longrightarrow$ $^{50}$Ca&\quad\quad33.768  \quad31.906\quad-1.862  \\
$^{236}$Np $\longrightarrow$ $^{29}$Mg&\quad\quad26.044  \quad26.701\quad 0.657  &&$^{251}$Cf $\longrightarrow$ $^{46}$Ar&\quad\quad30.023  \quad28.044\quad-1.979  \\
$^{236}$Pu $\longrightarrow$ $^{28}$Mg&\quad\quad21.180  \quad21.107\quad-0.073  &&$^{252}$Cf $\longrightarrow$ $^{46}$Ar&\quad\quad29.450  \quad27.371\quad-2.079  \\
$^{236}$Pu $\longrightarrow$ $^{30}$Mg&\quad\quad27.608  \quad28.051\quad 0.444  &&$^{252}$Md $\longrightarrow$ $^{46}$Ca&\quad\quad28.135  \quad25.713\quad-2.422  \\
\hline
\end{tabular}
\newpage
\begin{tabular}{|lll|   |llll|}
\hline
cluster decay &\quad\quad ASAF \quad TD. \qquad $\Delta$ && cluster decay &\quad\quad ASAF \quad TD. \qquad $\Delta$\\
\hline
$^{237}$Np $\longrightarrow$ $^{32}$Si&\quad\quad27.691  \quad28.824\quad 1.133  &&$^{290}$Lv  $\longrightarrow$ $^{28}$Mg &\quad\quad28.151  \quad31.051\quad 2.900\\
$^{237}$Pu $\longrightarrow$ $^{30}$Mg&\quad\quad26.473  \quad26.752\quad 0.279  &&$^{294}$118  $\longrightarrow$ $^{50}$Ca&\quad\quad30.932  \quad28.049\quad-2.883\\
$^{254}$Md $\longrightarrow$ $^{48}$Ca&\quad\quad26.229  \quad22.387\quad-3.843  &&$^{252}$No $\longrightarrow$ $^{50}$Ca  &\quad\quad30.370  \quad26.912\quad-3.458   \\
$^{254}$No $\longrightarrow$ $^{48}$Ca&\quad\quad25.771  \quad21.616\quad-4.155  &&$^{253}$No $\longrightarrow$ $^{50}$Ca  &\quad\quad29.953  \quad26.411\quad-3.542   \\
$^{255}$No $\longrightarrow$ $^{48}$Ca&\quad\quad25.129  \quad20.802\quad-4.328  &&$^{294}$118  $\longrightarrow$ $^{48}$Ca&\quad\quad28.298  \quad25.294\quad-3.004 \\
$^{256}$No $\longrightarrow$ $^{48}$Ca&\quad\quad24.856  \quad20.481\quad-4.374  &&$^{294}$118  $\longrightarrow$ $^{34}$Si&\quad\quad28.340  \quad29.541\quad 1.202 \\
$^{257}$No $\longrightarrow$ $^{48}$Ca&\quad\quad25.982  \quad22.075\quad-3.907  &&$^{294}$118  $\longrightarrow$ $^{28}$Mg&\quad\quad26.362  \quad28.647\quad 2.285 \\
$^{258}$Md $\longrightarrow$ $^{50}$Ca&\quad\quad28.580  \quad25.039\quad-3.540  &&$^{296}$120  $\longrightarrow$ $^{50}$Ca&\quad\quad29.703  \quad26.144\quad-3.559 \\
$^{258}$No $\longrightarrow$ $^{50}$Ca&\quad\quad26.552  \quad22.117\quad-4.435  &&$^{296}$120  $\longrightarrow$ $^{48}$Ca&\quad\quad26.588  \quad22.688\quad-3.900 \\
$^{258}$Rf $\longrightarrow$ $^{49}$Ca&\quad\quad27.898  \quad24.045\quad-3.854  &&$^{296}$120  $\longrightarrow$ $^{34}$Si&\quad\quad26.225  \quad26.588\quad 0.362 \\
$^{258}$Rf $\longrightarrow$ $^{50}$Ti&\quad\quad26.451  \quad21.964\quad-4.488  &&$^{296}$120  $\longrightarrow$ $^{28}$Mg&\quad\quad23.811  \quad25.164\quad 1.353 \\
$^{258}$Rf $\longrightarrow$ $^{52}$Ti&\quad\quad27.570  \quad22.681\quad-4.889  &&$^{296}$120  $\longrightarrow$ $^{22}$Ne&\quad\quad21.380  \quad23.695\quad 2.315 \\
$^{258}$Rf $\longrightarrow$ $^{54}$Ti&\quad\quad30.359  \quad25.722\quad-4.636  && \\
$^{259}$No $\longrightarrow$ $^{50}$Ca&\quad\quad27.544  \quad23.526\quad-4.019  && \\
$^{261}$No $\longrightarrow$ $^{48}$Ca&\quad\quad30.204  \quad28.182\quad-2.022  && \\
$^{262}$No $\longrightarrow$ $^{50}$Ca&\quad\quad30.314  \quad27.526\quad-2.788  && \\
$^{271}$Sg $\longrightarrow$ $^{50}$Ca&\quad\quad35.508  \quad34.739\quad-0.769  && \\
$^{271}$Sg $\longrightarrow$ $^{48}$Ca&\quad\quad35.301  \quad35.561\quad 0.260  && \\
$^{275}$Hs $\longrightarrow$ $^{50}$Ca&\quad\quad34.793  \quad33.712\quad-1.081  && \\
$^{275}$Hs $\longrightarrow$ $^{48}$Ca&\quad\quad33.969  \quad33.627\quad-0.342  && \\
$^{275}$Hs $\longrightarrow$ $^{34}$Si&\quad\quad32.585  \quad35.222\quad 2.637  && \\
$^{272}$Ds $\longrightarrow$ $^{50}$Ca&\quad\quad32.760  \quad30.290\quad-2.471  && \\
$^{272}$Ds $\longrightarrow$ $^{48}$Ca&\quad\quad30.114  \quad27.558\quad-2.556  && \\
$^{272}$Ds $\longrightarrow$ $^{34}$Si&\quad\quad29.647  \quad30.659\quad 1.012  && \\
$^{272}$Ds $\longrightarrow$ $^{28}$Mg&\quad\quad26.760  \quad28.470\quad 1.710  && \\
$^{272}$Ds $\longrightarrow$ $^{22}$Ne&\quad\quad26.238  \quad29.287\quad 3.049  && \\
$^{290}$Lv  $\longrightarrow$ $^{50}$Ca&\quad\quad32.079  \quad29.703\quad-2.375&& \\
$^{290}$Lv  $\longrightarrow$ $^{48}$Ca&\quad\quad29.710  \quad27.344\quad-2.367&& \\
$^{290}$Lv  $\longrightarrow$ $^{34}$Si&\quad\quad29.450  \quad31.000\quad 1.550&& \\
\hline
\end{tabular}
}

\end{document}